\documentclass[aps,preprint,amsmath,tightenlines,showpacs,showkeys]{revtex4}

\usepackage{bm}

\begin{document}

\def\US{\Upsilon}
\def\zb{Z^{\pm}_b}
\def\zbp{Z^{'\pm}_b}

\title{Sum rules for tetraquark decay coupling constants with broken SU(3)
symmetry}

\author{V.~Gupta}

\email[]{virendra@mda.cinvestav.mx}

\affiliation{
Departamento de F\'{\i}sica Aplicada.\\
Centro de Investigaci\'on y de Estudios Avanzados del IPN.\\
Unidad M\'erida.\\
A.P. 73, Cordemex.\\
M\'erida, Yucat\'an, 97310. MEXICO.
}

\author{G.~S\'anchez-Col\'on}

\email[]{gsanchez@mda.cinvestav.mx}

\affiliation{
Departamento de F\'{\i}sica Aplicada.\\
Centro de Investigaci\'on y de Estudios Avanzados del IPN.\\
Unidad M\'erida.\\
A.P. 73, Cordemex.\\
M\'erida, Yucat\'an, 97310. MEXICO.
}

\author{S.~Rajpoot}

\email[]{rajpoot@csulb.edu}

\affiliation{
Department of Physics \& Astronomy.\\
California State University, Long Beach.\\
Long Beach, CA 90840. USA.
}

\date{\today}

\begin{abstract}

We give sum rules for tetraquark decay coupling constants, taking into
account the SU(3) symmetry breaking interactions to first order.

\end{abstract}

\pacs{14.40.Rt, 13.25.Jx}

\keywords{tetraquarks, sum rule}

\maketitle

\section{\label{introduction}Introduction}

Recently, two hidden-bottom charged meson resonances, referred to as $Z_b(10610)$
and $Z_b(10650)$, were observed by the Belle Collaboration~\cite{belle12} as two
narrow resonance structures in the mass spectra of $\pi^{\pm}\Upsilon(nS)$
($n=1,2,3$) and $\pi^{\pm}h_b(mP)$ ($m=1,2$). They are produced in association
with a single charged pion in $\Upsilon(5S)$ decays with the following values of
mass and width, averaged over the five final states:

\[
M[Z_b(10610)]=(10607.2\pm 2.0){\rm MeV}/c^2,\quad
\Gamma[Z_b(10610)]=(18.4\pm 2.4){\rm MeV},
\]

\begin{equation}
M[Z_b(10650)]=(10652.2\pm 1.5){\rm MeV}/c^2,\quad
\Gamma[Z_b(10650)]=(11.5\pm 2.2){\rm MeV}.
\end{equation}

\noindent
Two main theoretical structure assignments are proposed for these
hidden-bottom meson resonances. One is based on molecular
excitations~\cite{voloshin1,voloshin2}, the molecules being two meson bound
state with a very small binding energy, and the other one is the tetra-quark
interpretation~\cite{Guo,Richard} using the analogy similar to the charm
sector.

The $J^P$ of both, $Z_b(10610)$ and $Z_b(10650)$, is determined to be $1^+$.
This is consistent with their decay into a $1^-$-state ($\Upsilon(nS)$) and a
$0^-$-meson in a relative $s$-state. In this note we suggest that these new
states are tetraquarks made of $(b\bar{b}\,q\bar{q})$ with $q=u,d,s$. We work
out the consequences of broken SU(3) symmetry to derive sum rules for
tetraquark decay coupling constants.

\section{\label{notation}Notation}

Let us consider a general tetraquark $T(Q\bar{Q}\,q\bar{q}))$ with $Q$ ($=b\
{\rm or}\ c$) quark a SU(3) flavour singlet and $q\bar{q}$ made of the light
quarks $q$ ($=u,d,\ {\rm and}\ s$). One natural decay mode of the tetraquark
$T(Q\bar{Q}\,q\bar{q})$ would be

\begin{equation}
T(Q\bar{Q}\,q\bar{q})\to H(Q\bar{Q}) +  M(q\bar{q}),
\label{tetraquark}
\end{equation}

\noindent
where $H(Q\bar{Q})$ is a heavy meson since $Q=b\ {\rm or}\ c$ and
$M(q\bar{q})$ could be the SU(3) flavour $0^-$ pseudoscalar octet \textbf{P},

\begin{eqnarray}
\bf{P}&=&\left(\begin{array}{ccc}
\frac{1}{\sqrt{2}}\pi^{0}+\frac{1}{\sqrt{6}}\eta&\pi^{+}&K^{+}\\
\pi^{-}&-\frac{1}{\sqrt{2}}\pi^{0}+\frac{1}{\sqrt{6}}\eta&K^{0}\\
K^- &\bar{K}^{0}&-\frac{2}{\sqrt{6}}\eta
\end{array}\right), \label{pseudo}
\end{eqnarray}

\noindent
or the $1^-$ vector octet \textbf{V},

\begin{eqnarray}
\bf{V}&=&\left(\begin{array}{ccc}
\frac{1}{\sqrt{2}}\rho^{0}+\frac{1}{\sqrt{2}}\omega&\rho^{+}&K^{*+}\\
\rho^{-}&-\frac{1}{\sqrt{2}}\rho^{0}+\frac{1}{\sqrt{2}}\omega&K^{*0}\\
K^{*-} &\bar{K}^{*0}&\phi
\end{array}\right).\label{vector}
\end{eqnarray}

\noindent
Furthermore, one expects that $H(Q\bar{Q})$ and $M(q\bar{q})$ in the final
state have the same $J^P$ they have in the tetraquark. This is the simplest
possibility.

\section{\label{sumrule}Sum rules}

If one assumes that the decay interaction is flavour SU(3) singlet, then all the
eight decays will have the same coupling constant $G_0$. However if one
considers a first order breaking interaction transforming like $\lambda_8$ (or
$T^3_3$) which conserves isospin invariance then this will give rise to two
SU(3) breaking coupling constants, $G_F$ and $G_D$, which arise from the
antisymmetric and symmetric octets resulting from the breaking interaction and
the $M(q\bar{q})$ octet in the final state. Thus, the eight decays will be
given in terms of three coupling constants. However, since isospin invariance
is present, there will be four independent coupling constants namely, when the
$M(q\bar{q})$ meson is ($K^+$, $K^0$), ($\bar{K}^0$, $K^-$), ($\pi^+$, $\pi^0$,
$\pi^-$), and $s\bar{s}$. These four coupling constants will be linear
combinations of $G_0$, $G_F$, and $G_D$, and hence, we expect a sum rule among
the four coupling constants of the observable
decays~\cite{gupta13564,gupta13664,gupta66}.

Since the SU(3) calculation is analogous to the calculation of the
meson masses with $\lambda_8$ breaking, one obtains a coupling constants
sum rule analogous to the Gell-Mann-Okubo mass formula for
mesons~\cite{gellmann61,okubo62}. Thus, if our interpretation of the observed
particles as tetraquarks is right we expect,

\begin{eqnarray}
\lefteqn{2\,G\left[T(Q\bar{Q}\,K)\to H(Q\bar{Q}) + K\right] +
2\,G\left[T(Q\bar{Q}\,K)\to H(Q\bar{Q}) + \bar{K}\right] = } \nonumber\\
 & & G\left[T(Q\bar{Q}\,\pi)\to H(Q\bar{Q}) + \pi\right] +
3\,G\left[T(Q\bar{Q}\,s\bar{s})\to H(Q\bar{Q}) + M(s\bar{s})\right].
\label{sumrulet}
\end{eqnarray}

\noindent
Here $K=K^+$ or $K^0$, $\bar{K}=\bar{K}^0$ or $K^-$, $\pi=\pi^+$,
$\pi^-$, or $\pi^0$. The heavy quark $Q$ is a SU(3) singlet and could be
either $b$ or $c$ quark.

The above formula for decay coupling constants is also valid for tetraquarks
$T(Q\bar{Q}'\,M(q\bar{q}))$ where $Q=b$ (or $c$) with $Q'=c$ (or $b$).

As noted earlier, the light quark octet $M(q\bar{q})$ could be the $1^-$ vector
octet ($\rho$, $K^*$, etc.). In this case, the sum rules would be valid with
the obvious changes $\pi\to\rho$, $K\to K^*$, etc. The sum rules contain
$M(s\bar{s})$ and represents $\eta$, $\eta'$, $\phi$ or $\omega$ as
appropriate.

Applied to specific decay modes of the tetraquarks, the sum rule translates into
the following sum rules,

\begin{table}[ht]
\centering
\begin{tabular}{c}
$2 G_{K^+} + 2G_{\bar{K}^0}=G_{\pi^+}+3G_{\eta}$,\\
$2 G_{K^-} + 2G_{K^0}=G_{\pi^-}+3G_{\eta} $,\\
$2 G_{K^0} + 2G_{\bar{K}^0}=G_{\pi^0}+3G_{\eta}$,\\
$2 G_{K^+} + 2G_{\bar{K}^0}=G_{\pi^+}+3G_{\eta'}$,\\
$2 G_{K^-} + 2G_{K^0}=G_{\pi^-}+3G_{\eta'}$, \\
$2 G_{K^0} + 2G_{\bar{K}^0}=G_{\pi^0}+3G_{\eta'}$, \\
\end{tabular}
\end{table}

\noindent
where we have abbreviated the couplings as $G\left[T(Q\bar{Q}\,K^+)\to
H(Q\bar{Q}) + K^+ \right] \equiv G_{K^+} $, and similarly for the other
couplings.

As is well known, both for pseudoscalar and vector mesons, $s\bar{s}$ is a
combination of the SU(3) singlet $\eta_1$ and the eighth component $\eta_8$.
This mixing must be taken into account in testing our coupling constants sum
rule. Specifically, in terms of the SU(3) flavor states $\eta_1$ and $\eta_8$,

\begin{equation}
s\bar{s}=\frac{1}{\sqrt{3}}\,\eta_1 -\frac{2}{\sqrt{6}}\,\eta_8,
\label{ss}
\end{equation}

\noindent
where

\begin{equation}
\eta_1=\frac{1}{\sqrt{3}}\,(u\bar{u} + d\bar{d} +
s\bar{s})
\quad {\rm and} \quad
\eta_8=\frac{1}{\sqrt{6}}\,(u\bar{u} + d\bar{d} -
2s\bar{s}).
\label{eta1eta8}
\end{equation}

\noindent
For $J^P=0^-$, $\eta(548)=\eta_8\cos{\theta}-\eta_1\sin{\theta}$ and
$\eta'(958)=\eta_8\sin{\theta}+\eta_1\cos{\theta}$, with
$\theta=-11.5^{\circ}$. For $J^P=1^-$, $\eta(548)\to\phi(1020)$ and
$\eta'(958)\to\omega(782)$, with $\theta=38.7^{\circ}$~\cite{pdg10}.

The various couplings entering the sum rules can be extracted from the
appropriate decay widths. For the $Z_b(10610)$ and $Z_b(10650)$ tetraquark
decays into two particles, a $1^-$-state ($\Upsilon(nS)$) and a $0^-$-meson, in
a relative $s$-state the width is given by

\begin{equation}
\Gamma=\frac{1}{8\pi}\frac{k}{M^2}|A|^2,
\end{equation}

\noindent
where $M$ is the tetraquark mass, $k$ is the momentum of a decay
particle and $A$ is the transition amplitude. For $s$-wave decay we take
$A=GM$, so that

\begin{equation}
G^2= \frac{8\pi}{k}\Gamma.
\end{equation}

\noindent
Numerical results for $G\left[Z_b(10610)\to \Upsilon(nS) + \pi^{+}
\right]$ and $G\left[Z_b(10650)\to \Upsilon(nS) + \pi^{+} \right]$ ($n=1,2,3$)
are presented in Table~\ref{tablaI}~\cite{pdg10}. The sum rules can hopefully
be verified when data for other decay modes of $Z_b$-tetraquarks are available.

\section{\label{conclusions}Conclusions}

In anticipation of more solid data on tetraquarks in the future~\cite{review},
we have derived sum rules for tetraquark decay coupling constants assuming
broken SU(3) symmetry only up to the first order. Our sum rules,
Eqs.~(\ref{sumrulet}), rely on the assumption that the heavy quarks are
spectator singlets. It will be interesting to check  these sum rules  when the
tetraquark spectroscopy gets better defined in the future.

\begin{acknowledgments}

V.~Gupta and G.~S\'anchez-Col\'on would like to thank CONACyT (M\'exico) for
partial support. The work of S.~Rajpoot was supported by DOE Grant \#:
{DE-FG02-10ER41693}.

\end{acknowledgments}

\clearpage

\begin{table}

\caption{Numerical results for $Z_b(10610)$ and $Z_b(10650)$ parameters obtained
from $Z_b(10610)\to \Upsilon(nS) + \pi^{+}$ and $Z_b(10650)\to \Upsilon(nS) +
\pi^{+}$ ($n=1,2,3$) analysis. \label{tablaI}}

\begin{ruledtabular}

\begin{tabular}{lccc}

 & $\Upsilon(1S)\pi^{+}$ & $\Upsilon(2S)\pi^{+}$ & $\Upsilon(3S)\pi^{+}$ \\

\hline

$M[Z_b(10610)]$ (MeV) & $10611\pm 4$ & $10609\pm 2$ & $10608\pm 2$
\\

$\Gamma[Z_b(10610)]$ (MeV) & $22.3\pm 7.7$ & $24.2\pm 3.1$ &
$17.6\pm 3.0$ \\

$k$ (MeV) & $1080.3\pm 3.6$ & $553.2\pm 2.0$ &
$208.3\pm 2.4$ \\

$G^2$ & $0.52\pm 0.18$ & $1.10\pm 0.14$ &
$2.12\pm 0.36$ \\

\hline

$M[Z_b(10650)]$ (MeV) & $10657\pm 6$ & $10651\pm 2$ & $10652\pm 1$
\\

$\Gamma[Z_b(10650)]$ (MeV) & $16.3\pm 9.8$ & $13.3\pm 3.3$ &
$8.4\pm 2.0$ \\

$k$ (MeV) & $1121.8\pm 5.4$ & $594.0\pm 2.0$ &
$258.3\pm 1.2$ \\

$G^2$ & $0.37\pm 0.22$ & $0.56\pm 0.14$ &
$0.82\pm 0.19$ \\

\end{tabular}

\end{ruledtabular}

\end{table}


\begin{thebibliography}{}

\bibitem{belle12}

A.~Bondar {\it et al.} (Belle Collaboration), Phys.\ Rev.\
Lett.\ {\bf 108} (2012) 122001.

\bibitem{voloshin1}
  A.~E.~Bondar, A.~Garmash, A.~I.~Milstein, R.~Mizuk and M.~B.~Voloshin,
  Phys.\ Rev.\  D {\bf 84} (2011) 054010.

\bibitem{voloshin2}
  M.~B.~Voloshin,
  Phys.\ Rev.\  D {\bf 84} (2011) 031502.

\bibitem{Guo}
  T.~Guo, L.~Cao, M.~Z.~Zhou and H.~Chen,
  ``The possible candidates of tetraquark : $Z_b(10610)$ and $Z_b(10650)$'',
 [arXiv:1106.2284 [hep-ph]].

\bibitem{Richard}
  F.~S.~Navarra, M.~Nielsen, J.~-M.~Richard,
  ``Exotic Charmonium and Bottomonium-like Resonances'',
  [arXiv:1108.1230 [hep-ph]].

\bibitem{gupta13564}

V.~Gupta and V.~Singh, Phys.\ Rev.\ {\bf 135} (1964) 1442.

\bibitem{gupta13664}

V.~Gupta and V.~Singh, Phys.\ Rev.\ {\bf 136} (1964) 782.

\bibitem{gupta66}

V.~Gupta, Il Nvo.\ Cim.\ {\bf 42} (1966) 737.

\bibitem{gellmann61}

M.~Gell-Mann, {\it ``The Eightfold Way: A Theory of Strong
Interaction Symmetry"}, California Institute of Technology Synchrotron
Laboratory Report CTSL-20 (1961), unpublished.

\bibitem{okubo62}

S.~Okubo, Prog.\ Theor.\ Phys.\ {\bf 27} (1962) 949.

\bibitem{pdg10}

K.~Nakamura {\it et al.} (Particle Data Group), J.\ Phys.\ G.\ {\bf 37}
(2010) 075021.

\bibitem{review}

S. Eidelman {\it et al.},``Developments in heavy Quark Spectroscopy",
[arXiv:1205:4189 [hep-ex]].

\end{thebibliography}
\end{document}